\documentclass[prl,comment,twocolumn,showpacs,preprintnumbers,amsmath,amssymb]{revtex4}
\usepackage{graphicx}
\begin{document}

\noindent {\bf  Comment on ``Dephasing Times in Quantum Dots
 due to Elastic LO Phonon-Carrier Collisions''}\\

In a Letter~\cite{Uskov00} Uskov \emph{et al.} proposed a
second-order elastic interaction between quantum dot (QD) charge
carriers and longitudinal optical (LO) phonons as an intrinsic
mechanism for dephasing in QDs. It was claimed in
Ref.~\cite{Uskov00} that this coupling may lead to dephasing times
of the order of 200\,fs at $T=300$\,K and to a Gaussian absorption
lineshape.

In the present Comment we show that this result by Uskov \emph{et
al.} is incorrect. We demonstrate that (i) the level-diagonal
quadratic coupling to dispersionless LO phonons leads in QDs to
\emph{no dephasing} at all, and (ii) the spectrum consists
exclusively of discrete lines.

In order to make things more clear, we disregard completely the
linear coupling $\hat{H}^{(1)}$ which is indeed zero if the
electron and hole confined functions are identical, as done in
Ref.~\cite{Uskov00}, and we neglect any phonon damping. We use
$\hat{H}^{(2)}=-\sum_{{\bf q},{\bf q}'} F({\bf q},{\bf q}')
(a_{\bf q}+a^\dagger_{-\bf q}) (a_{-{\bf q}'}+a^\dagger_{{\bf
q}'})$ as quadratic coupling with a factorizable function $F({\bf
q},{\bf q}')$ (assuming only one electron and one hole excited
state). Then, as shown in~\cite{Muljarov04}, the linear optical
polarization can be given \emph{exactly} by $P(t)=\exp\{K(t)\}$
with the cumulant $K(t)$ being an infinite sum of all-order
diagrams,
\begin{figure}[b]
\includegraphics*[angle=-90,width=8.6cm]{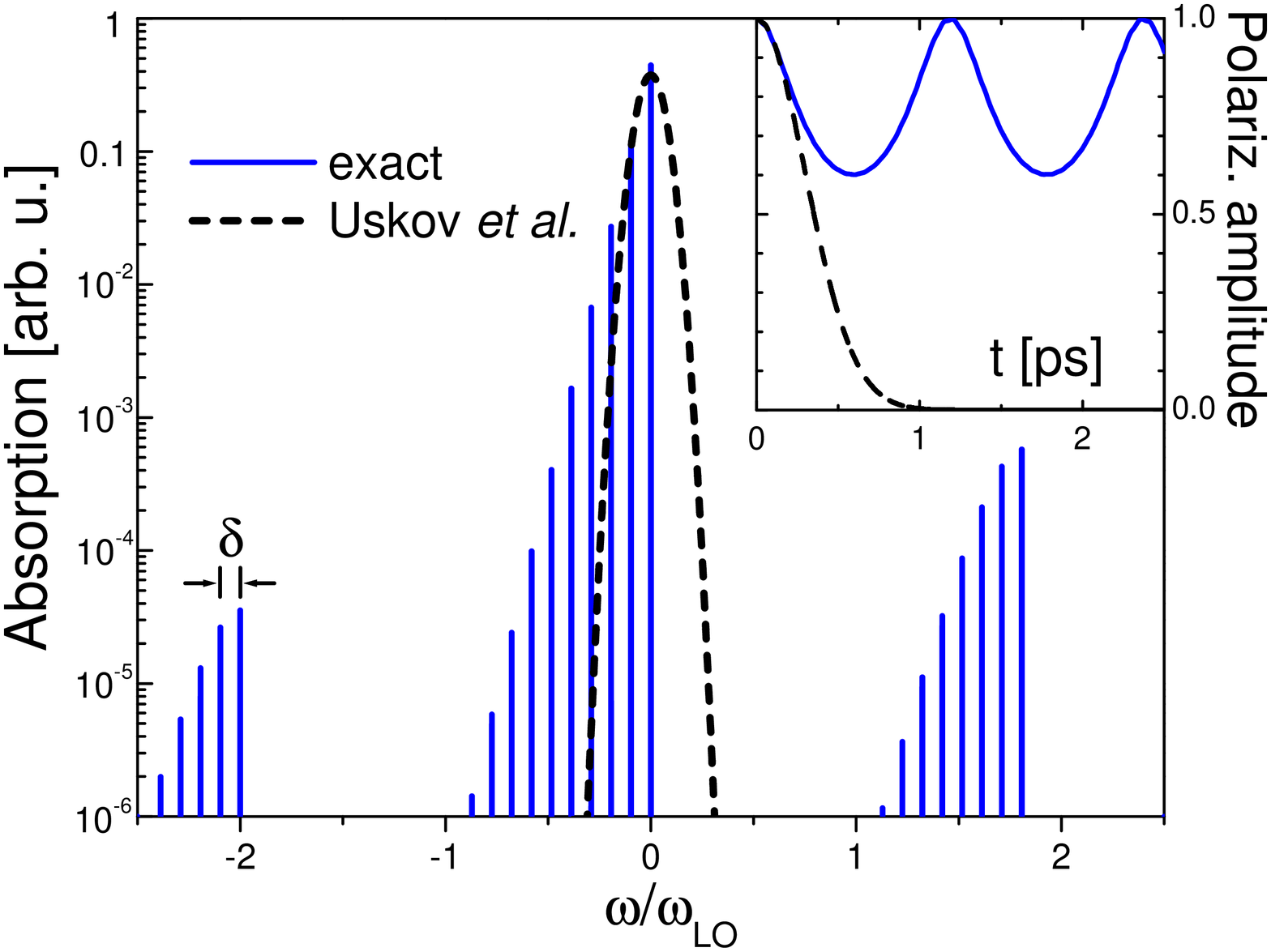}
\caption{Absorption spectrum and polarization amplitude (inset)
calculated exactly, using Eq.~(\ref{cumulant}), and up to second
order in the cumulant, Eq.~(\ref{Uskov}), as done in
Ref.~\cite{Uskov00}. Parameter values are the same as in
Ref.~\cite{Uskov00}: $T=300$\,K, $\hbar\omega_{\rm LO}=36$\,meV,
$\hbar\sigma=\hbar\varkappa\sqrt{\bar{n}(\bar{n}+1)}=2.2$\,meV. }
\end{figure}
\begin{equation}
K(t)=-\frac{1}{2}\sum_{j=1}^\infty\ln[1-i\varkappa\lambda_j(t)],
\label{cumulant}
\end{equation}
where $\varkappa=(2/\hbar)\sum_{\bf q} F({\bf q},{\bf q})$.
$\lambda_j$ are all possible eigenvalues of the Fredholm problem,
Eq.~(10) of Ref.~\cite{Muljarov04}. For the present case of
dispersionless LO phonons, the eigenvalues can be obtained
\mbox{straightforwardly:} \mbox{$\omega_{\rm
LO}\,\lambda_j=2i/(s_j^2-1)$,} where $s_j$ are the solutions of
\begin{equation}
\tan\left(\frac{s\,\omega_{\rm
LO}\,t+\pi}{2}\right)=s\,\tan\left(\frac{\omega_{\rm LO}\,t+\pi
+i\beta}{2}\right)
 \label{roots}
\end{equation}
for even eigenfunctions ($\beta=\hbar\omega_{\rm LO}/k_BT$). For
the odd solutions, drop $\pi$ in Eq.~(\ref{roots}). The inclusion
of two degenerate excited states for both carriers (as done in
Ref.~\cite{Uskov00}) will lead only to an additional factor of 2
in Eq.~(\ref{cumulant}).

Uskov \emph{et al.} calculated the same cumulant only up to second
order (see Eq.~(12) in Ref.~\cite{Uskov00}), which would read
\begin{equation}
K'(t)=\sum_{j=1}^\infty \left[ \frac{i\varkappa\lambda_j(t)}{2} -
\frac{\varkappa^2\lambda^2_j(t)}{4}\right].
 \label{Uskov}
\end{equation}
Then the amplitude of the polarization $P'(t)=\exp\{K'(t)\}$ is a
rapidly decaying (nearly Gauss) function (dashed curve in the
inset of Fig.1), while the exact polarization is almost perfectly
periodic (full curve). In order to check that the full solution
indeed does not lead to any dephasing we have also diagonalized
exactly the Hamiltonian in the spirit of Ref.~\cite{Stauber00-05}.
The only phonon mode which couples to the electronic system
carries a renormalized frequency $\tilde{\omega}_{\rm
LO}=\sqrt{\omega^2_{\rm LO}-2\varkappa\omega_{\rm LO}}$, and the
absorption (Fig.1) represents a set of discrete delta-lines,
$\alpha(\omega)=\sum_{n,m=0}^\infty\alpha_{nm}\delta(\omega-n\tilde{\omega}_{\rm
LO}+m\omega_{\rm LO})$, in agreement with Ref.~\cite{Stauber00-05}
where the full problem of a few electronic and hole levels in a QD
coupled to LO-phonons is solved exactly. Note the fine structure
of satellites separated by $\delta=|\omega_{\rm
LO}-\tilde{\omega}_{\rm LO}|$, in addition to the standard
two-phonon satellites asymmetrically positioned around the
zero-phonon line. In contrast, the Gauss-like spectrum found by
Uskov \emph{et al.} (dashed curve) has a finite linewidth.

In conclusion, we have shown that the finite dephasing times in
QDs derived in Ref.~\cite{Uskov00} are an artefact due to a
truncation of the infinite series in the cumulant. Dispersionless
LO phonons quadratically coupled to a single QD transition produce
no dephasing, but a rich line spectrum, in agreement with a more
general approach~\cite{Stauber00-05}. \vfill

\noindent E. A. Muljarov and R. Zimmermann

Institut f\"ur Physik

Humboldt-Universit\"at zu Berlin

Newtonstrasse 15, D-12489 Berlin, Germany\\

\noindent 27 September 2005


\itemize
\bibitem{Uskov00} A.\,V. Uskov, A.-P. Jauho, B. Tromborg, J. M\o rk, and R.
Lang, Phys. Rev. Lett. {\bf 85}, 1516 (2000).
\vspace{-2.7mm}
\bibitem{Muljarov04} E.\,A. Muljarov and R. Zimmermann, Phys. Rev. Lett. {\bf 93}, 237401 (2004).
\vspace{-2.7mm}
\bibitem{Stauber00-05} T.\,Stauber,\,R.\,Zimmermann,\,and H.\,Castella,
Phys. Rev. B {\bf 62}, 7336 (2000); arXiv:cond-mat/0506408.
\end{document}